\documentstyle[aps,floats,preprint,epsfig]{revtex}
\tighten
\begin{document}
\begin{titlepage}
\thispagestyle{empty}
%\preprint{FIS-UI-TH}
\title{Kaon photoproduction on the nucleon : 
  Contributions of kaon-hyperon final states to the magnetic moment of the 
  nucleon}
\author{S. Sumowidagdo and T. Mart} 
\address{Jurusan Fisika, FMIPA, Universitas Indonesia, Depok 16424, Indonesia}
%\date{}
\maketitle
\begin{abstract}
By using the Gerasimov-Drell-Hearn (GDH) sum rule and an isobaric model of
kaon photoproduction, we calculate contributions of kaon-hyperon final states 
to the magnetic moment of the proton and the neutron.  We find that the 
contributions are small.  The approximation of $\sigma_{TT'}$ by $\sigma_{T}$ 
clearly overestimates the value of the GDH integral.  We find a smaller upper
bound  for the contributions of kaon-hyperon final states to the proton's
anomalous magnetic moment in kaon photoproduction, and a positive contribution
for the square of the neutron's magnetic moment. \\
PACS number(s): 13.60.Le,11.55.Hx, 13.40.Em, 14.20.Dh
\end{abstract}

\end{titlepage}

The internal structure of the nucleon is still an interesting topic of
investigations nowadays.  The existence of this structure is responsible for
the ground state properties of the nucleon, such as hadronic and
electromagnetic form factors and the anomalous magnetic moment.  At higher
energies this finite internal structure yields a series of resonances in the
mass region of 1 $-$ 2 GeV.  It was then found that the nucleon's ground state
properties and the nucleon's resonance spectra are not all independent
phenomena;  they are related by a number of sum rules \cite{dgdh}. 

One of these sum rules is the Gerasimov-Drell-Hearn (GDH) sum rule, which
connects the nucleon's magnetic moments and the helicity structures in the
resonance region.  Although the GDH sum rule was proposed more than 30 years
ago, no direct experiment had been performed to investigate whether or not
the sum rule converges. However, with the advent of the new high-intensity and 
continuous-electron-beam accelerator, accurate measurements of
the contribution to the GDH integral from individual final states are made
possible. 

Previously, Hammer, Drechsel, and Mart (HDM)  suggested that by using the
Gerasimov-Drell-Hearn sum rule it is possible to estimate strange
contributions to the magnetic moments of the proton \cite{hdm}.  They used
experimental data and an isobaric model for the photoproduction of $\eta$,
$\phi$, as well as $K$ mesons, in order to estimate the transversely
unpolarized total cross section $\sigma_T$ and, therefore, to calculate the
upper bounds of strange contributions to the anomalous magnetic moment of the
proton.  It is the purpose of this Brief Report to update the contributions of
kaon-hyperon final states, by means of the latest isobaric model which fits
all available experimental data, including the recent data from \textsc{SAPHIR}
\cite{saphir}. 

The GDH sum rule \cite{gdh} (for a review see Ref. \cite{dgdh}) relates the
anomalous magnetic moment of the nucleon $\kappa_N$  to the difference of its
polarized total photoabsorption cross section
 \begin{equation}
  \label{gdhsr}
  -\frac{\kappa_N^2}{4}~=~\frac{m_N^2}{8\pi^2\alpha}\int_0^\infty 
  \frac{d\nu}{\nu} [ \sigma_{1/2}(\nu)-\sigma_{3/2}(\nu)],
\end{equation}
where $\sigma_{3/2}$ and $\sigma_{1/2}$ denote the cross sections for the
possible combinations of spins of the nucleon and photon (i.e., $\sigma_{3/2}$
for total spin = $\case{3}{2}$ and $\sigma_{1/2}$ for total spin =
$\case{1}{2}$), $\alpha$ is the fine structure constant, $\nu$ is the photon 
energy in the laboratory frame, and $m_N$ the mass of the nucleon.  The 
derivation of GDH sum rule is based on general principles: Lorentz and gauge
invariance, crossing symmetry, causality, and unitarity.  The only assumption
in deriving Eq. (\ref{gdhsr}) is that the scattering amplitude goes to zero
for the limit $|\nu| \rightarrow \infty$, thus there is no subtraction
hypothesis \cite{arg}.
 
In photoproduction processes, however, the spin-dependent cross section is
related to the total cross sections by 
\begin{eqnarray}
  \label{sf}
  \sigma_{T} &=& \frac{\sigma_{3/2} + \sigma_{1/2}}{2},\\
  \sigma_{TT'} &=& \frac{\sigma_{3/2} - \sigma_{1/2}}{2}.
\end{eqnarray}
The first cross section can be measured using unpolarized real photons while
the second can be measured with longitudinally polarized electrons and
polarized nucleon targets or hyperon recoils.  Experimentally, the latter must
be done using electroproduction, i.e. virtual photons. Nevertheless, the
momentum transfer of the electrons ($Q^2$) can be minimized close to the 
photon point.  Numerically, $\sigma_{TT'}$ can be calculated by using
photoproduction, since Eq. (\ref{gdhsr}) needs $Q^2=0$ and
$\sigma_{TT'}=\sigma_{TT'}(F_1,F_2,F_3,F_4)$, where the $F_i$'s are the CGLN 
amplitudes for real photons \cite{knochlein}. 

Unlike the calculation in the previous paper, here we use both
\begin{eqnarray}
  \label{calc1}
  {\kappa_N^2} &=& \frac{m_N^2}{\pi^2\alpha}
  \int_0^{\nu_{\rm max}} \frac{d\nu}{\nu}\; \sigma_{TT'}
\end{eqnarray}
and
\begin{eqnarray}
  \label{calc2}
  {\kappa_N^2} &\lesssim&\frac{m_N^2}{\pi^2\alpha}
  \int_0^{\nu_{\rm max}} \frac{d\nu}{\nu}\; \sigma_{T}
\end{eqnarray}
where the GDH Integral is already saturated at $\nu_{\rm max} \approx$ 2 GeV 
\cite{burkert}, in order to measure the deviations of the approximation made
by the previous work from the expected values. This was not done in the
previous work since experimental data for kaon photoproduction were very
scarce at that time, especially for the $\gamma p \rightarrow K^0 \Sigma^+$
channel, thus predictions of $\sigma_{TT'}$ were somewhat unreliable.

We use the latest and modern elementary operator \cite{model}, which was 
guided by recent coupled-channel results \cite{feuster} and includes 
the newest data \cite{saphir}. The model consists of a tree-level amplitude
that reproduces all available $K^+ \Lambda$, $K^+ \Sigma^0$ and $K^0 \Sigma^+$
photoproduction observables and thus provides an effective parametrization
of these processes. The background terms contain the standard
$s$-, $u$-, and $t$-channel contributions along with a contact
term that was required to restore gauge invariance after hadronic
form factors had been introduced \cite{hbmf}.  This model includes the three
nucleon resonances that have been found in the coupled-channels approach to
decay into the $K \Lambda$ channel, the $S_{11}$(1650), $P_{11}$(1710), and
$P_{13}(1720)$. For $K \Sigma$ production further contributions from the
$S_{31}$(1900) and $P_{31}$(1910) $\Delta$ resonances were added. 

\begin{figure}
  \begin{center}  
    \leavevmode
    \epsfig{file=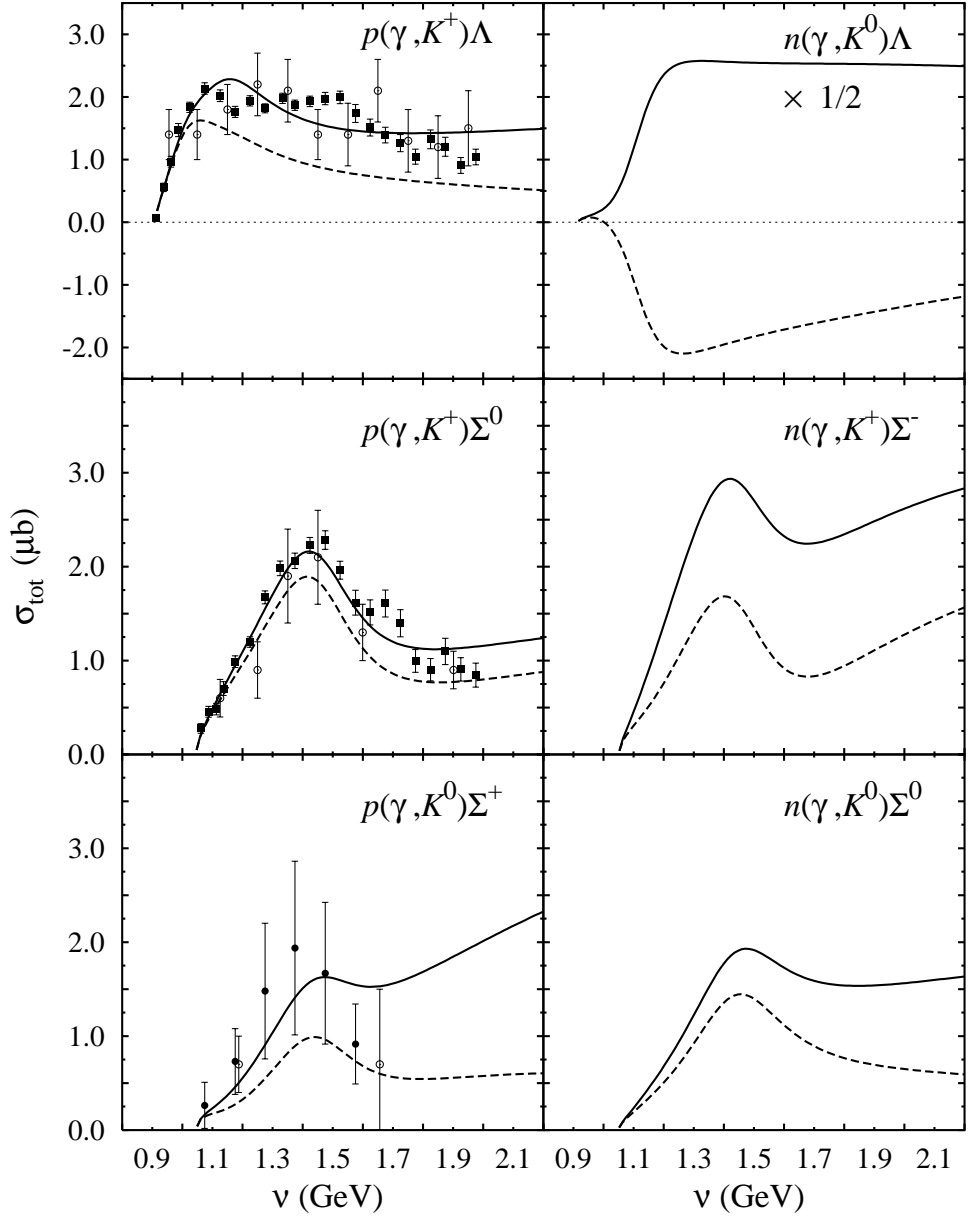,width=140mm}
    \caption{\label{fig:sffig} Total cross sections $\sigma_{T}$ (solid lines)
    and $-\sigma_{TT'}$ (dotted lines) for the six isospin channels plotted as 
    a function of  photon laboratory energy $\nu$ (GeV).  The elementary model
    is from Ref. \protect\cite{model}, experimental data are taken from Ref.
    \protect\cite{saphir}, and references therein. The elementary model fits
    not only total cross section data shown in this figure, 
    but also differential cross section and polarization data (not shown).  
    Total cross sections for the $n(\gamma,K^0)\Lambda$ channel are scaled
    with a factor of $\case{1}{2}$.}  
  \end{center}
\end{figure}

In Fig. \ref{fig:sffig} we show the total cross sections $\sigma_T$
and  $-\sigma_{TT'}$ as a function of the photon laboratory energy $\nu$ for
the six isospin channels in kaon photoproduction. Since there are no
experimental data for productions on the neutron, we consider the three right
panels in Fig. \ref{fig:sffig} as predictions. Obviously, the model can
remarkably reproduce the experimental data for the productions on the proton.
In the former calculation, contribution from the $\gamma p \rightarrow K^0
\Sigma^+$ channel could not properly be calculated since previous elementary
models mostly overpredict $K^0 \Sigma^+$ total cross section by a factor of up
to 100 \cite{mbh}. With the new SAPHIR data available in three isospin
channels, the elementary model becomes more reliable to explain kaon
photoproduction on the proton and to predict the production on the neutron. 

The elementary model predicts negative sign for $\sigma_{TT'}$ (note that we
have plotted $-\sigma_{TT'}$), except for the $K^0\Lambda$ channel, where it
produces a negative sign for the GDH integral of the neutron, thus yielding
positive values for $\kappa^2$ of the neutron, albeit $\gamma n \rightarrow
K^+ \Sigma^-$and $\gamma n \rightarrow K^0 \Sigma^0$ channels show a different
behavior. 
\begin{table}[t]
    \caption{Numerical values for the contribution of kaon-hyperon final
      states to the square of proton's and neutron's 
      anomalous magnetic moments $\kappa_N^2(K)$. 
      Column (1) is obtained from Eq. (\ref{calc1}), while column (2) 
      is evaluated by using Eq. (\ref{calc2}).}
    \begin{center}
    \begin{tabular}{lrrlrr}
Channel &\multicolumn{2}{c}{$\kappa_p^2(K)$}&
Channel&\multicolumn{2}{c}{$\kappa_n^2(K)$}\\
\cline{2-3}\cline{5-6}
&(1)&(2)&&(1)&(2)\\
\tableline
$\gamma\; p \rightarrow K^+ \Lambda$&$-0.026$&$0.044$&$\gamma\; 
n \rightarrow K^0 \Lambda$&$0.075$&$0.110$\\
$\gamma\; p \rightarrow K^+ \Sigma^0$&$-0.024$&$0.030$&$\gamma\; 
n \rightarrow K^+ \Sigma^-$&$-0.025$&$0.050$\\
$\gamma\; p \rightarrow K^0 \Sigma^+$&$-0.013$&$0.031$&$\gamma\; 
n \rightarrow K^0 \Sigma^0$&$-0.019$&$0.031$\\
\hline
Total&$-0.063$&$0.105$&Total&$0.031$&$0.191$\\
    \end{tabular}
  \end{center}
    \label{tab:contrib}
\end{table}

In Table \ref{tab:contrib} we list the numerical values obtained both by 
Eqs. (\ref{calc1}) and (\ref{calc2}), using a cutoff energy where we found the
elementary model is still reliable.  It is found that the result is not
sensitive to the cutoff energy $\nu_{\rm max}$ around 2 GeV, i.e. there is
no significant change in the integral in the energy interval 1.8 $-$ 2.2 GeV,
especially in the case of photoproduction on the proton where the cross
sections show a convergence at higher energies.  From Table \ref{tab:contrib}
it is already obvious that replacing Eq. (\ref{calc1}) by Eq. (\ref{calc2})
would overestimates the value of the GDH Integral, especially since we know 
that $\sigma_T$  is positive definite, while $\sigma_{TT'}$ is not.  
We find that our present calculation yields a slightly different result for
$\gamma p \rightarrow K^+ \Lambda$ channel, but not in the $\gamma p
\rightarrow K^+ \Sigma^0$ channel, where previous work seems to overestimate
the present calculation.   

Should the contributions add up coherently, our calculation would yield 
values of $\kappa_p^2(K)= -0.063$ and $\kappa_n^2(K)= 0.031$, or
$|\kappa_p(K)|/\kappa_p \leq 0.14$ and $\kappa_n(K)/\kappa_n \leq 
0.094$.  This put even smaller values for the upper bound of the magnitude of
kaon-hyperon final states contributions to the proton's magnetic moment,
compared to the previous result of HDM, $\kappa^2_p(K)=-0.07$ \cite{hdm}.  An
interesting feature is that our calculation yields a positive value for
contributions to the  $\kappa^2_n(K)$, therefore increases the calculated
value of the GDH Integral for the neutron. 

In conclusion, we have refined the calculation of kaon-hyperon final states
contributions to the anomalous magnetic moment of the proton and predicted the
contributions for the case of the neutron, based on the experimental data of
kaon photoproduction and a modern isobaric model.  Experimental data for 
$\sigma_{T}$ in neutron's channels and $\sigma_{TT'}$ in all six isospin 
channels will strongly suppress the uncertainties in our calculation.
Therefore, future experimental proposals in MAMI, ELSA, TJNAF, or GRAAL should
address this topic as an important measurement in order to improve our
understanding of the nucleon's structure. 

It is a pleasure to acknowledge that this work was supported by the University
Research for Graduate Education (URGE) grant. 

%\newpage

\end{document}